\documentclass[preprintnumbers,amsmath,amssymbm,preprint]{revtex4}
\usepackage{epsfig}
\usepackage{graphicx}

\begin{document}
\title{Upper bound on the energies of the emitted Hawking quanta}
\author{Shahar Hod}
\address{The Ruppin Academic Center, Emeq Hefer 40250, Israel}
\address{ }
\address{The Hadassah Institute, Jerusalem 91010, Israel}
\date{\today}

\begin{abstract}
\ \ \ Using Thorne's hoop conjecture, it is argued that the energies
of the Hawking quanta emitted from canonical Schwarzschild black
holes are bounded from above by the simple quantum relation ${\cal
E}<{\cal E}_{\text{max}}={\hbar^{2/3}}/M^{1/3}$. In particular, it
is shown that, due to non-linear (self-gravity) effects of the
tunneling quanta, higher energy field modes are re-absorbed (rather
than escape to infinity) by the black hole.
\end{abstract}
\bigskip
\maketitle


\section{Introduction}

Using a semi-classical analysis, in which quantized fields are
linearly coupled to a classical black-hole spacetime, Hawking
\cite{Hawt} has revealed the physically important fact that, instead
of being black, black holes are intriguingly characterized by
filtered black-body emission spectra with well defined temperatures
\cite{Hawt,Bekt}.

In particular, for a spherically symmetric Schwarzschild black hole
of mass $M$ (and a corresponding horizon radius $r_{\text{H}}=2M$),
the semi-classical Hawking emission power per one degree of freedom
is given by the integral relation \cite{Page,Notelm}
\begin{equation}\label{Eq1}
P={{\hbar}\over{2\pi}}\sum_{l,m}\int_0^{\infty} {{\Gamma
\omega}\over{e^{\hbar\omega/T_{\text{BH}}}\pm1}}d\omega\  ,
\end{equation}
where the $-/+$ signs refer respectively to bosonic/femionic field
modes and the Bekenstein-Hawking temperature of the black-hole is
given by the functional relation \cite{Hawt,Bekt,Noteunit}
\begin{equation}\label{Eq2}
T_{\text{BH}}={{\hbar}\over{4\pi r_{\text{H}}}}\  .
\end{equation}

The frequency-dependent factors $\Gamma=\Gamma_{lm}(\omega)$, which
in most cases of physical interest should be computed numerically
\cite{Page}, quantify the linearized interaction of the emitted
field modes with the effective scattering potential of the curved
black-hole spacetime. Since these dimensionless greybody factors,
which characterize the composed black-hole-field system, scale as
$\Gamma(M\omega\ll1)\propto (M\omega)^n$ with $n>0$ in the
small-frequency $M\omega\ll1$ limit and as $\Gamma(M\omega\gg1)\to1$
in the large-frequency $M\omega\gg1$ limit \cite{Page}, the
black-hole emission spectrum has a characteristic peak at
$M\omega^{\text{peak}}=O(1)$ [see Eqs. (\ref{Eq1}) and (\ref{Eq2})].
This implies that the average energy of the emitted Hawking quanta
is characterized by the simple relation \cite{Notesim,Notelrg}
\begin{equation}\label{Eq3}
\bar{\cal E}\sim {{\hbar}\over{M}}\  .
\end{equation}

Formally, the Hawking integral relation (\ref{Eq1}) implies that the
black hole can, in principle (though with exponentially small
probabilities), emit field modes with arbitrarily large frequencies
(energies). However, energy conservation considerations enforce the
trivial upper bound
\begin{equation}\label{Eq4}
{\cal E}\leq M\
\end{equation}
on the energies of the emitted Hawking quanta.

The main goal of the present compact paper is to point out that a
{\it stronger} (and physically non-trivial) upper bound, which seems to
have gone unnoticed in previous analyzes of the black-hole
evaporation process, exists on the energies of the emitted Hawking
quanta. In particular, using Thorne's hoop conjecture \cite{Thorne},
we shall explicitly demonstrate below that there exists a critical
energy scale above which the emitted quanta are re-absorbed (rather
than escape to infinity) by the black hole.

\section{Hawking evaporation as a tunneling process, Thorne's hoop
conjecture, and an upper bound on the energies of the emitted
black-hole quanta}

Following the remarkably elegant analysis presented in \cite{Wilck},
we shall consider a physical scenario in which an Hawking quantum of
energy ${\cal E}$ tunnels out of a Schwarzschild black hole of mass
$M$. However, unlike the interesting analysis presented in
\cite{Wilck}, which treats the Hawking quanta as point-like
particles, in the present analysis we shall take into account the
finite spatial extension of the tunneling quanta. In particular, an
Hawking field mode of proper energy ${\cal E}$ is characterized by
the {\it finite} proper wavelength (Compton length)
\begin{equation}\label{Eq5}
\lambda=2\pi\hbar/{\cal E}\  .
\end{equation}
Immediately after the completion of the tunneling process of the
Hawking quantum through the black-hole horizon \cite{Wilck}, one has
a physical system which is composed of an Hawking field mode of
proper energy ${\cal E}=2\pi\hbar/\lambda$ which is gravitationally
coupled to a black hole of reduced mass $M-{\cal E(\lambda)}$ and a
corresponding horizon radius of $r^{\text{new}}_{\text{H}}=2[M-{\cal
E}(\lambda)]$.

In particular, following \cite{Wilck}, we assume that the tunneling
Hawking quantum of energy ${\cal E}$ materializes at
$r^{\text{new}}_{\text{H}}=2(M-{\cal E})$, just outside the final
position of the black-hole horizon. According to Thorne's hoop
conjecture \cite{Thorne}, the composed black-hole-Hawking-quantum
configuration will form an engulfing horizon (which would prevent
the Hawking quantum from escaping to infinity) if its circumference
radius $r_{\text{c}}$ is equal to (or smaller than) the
corresponding Schwarzschild (hoop) radius $r_{\text{hoop}}=2M$ of
the composed system [whose total energy (energy-at-infinity) is
still given, due to energy conservation considerations, by $(M-{\cal
E})+{\cal E}=M$].

The proper distance $l_{\text{hoop}}$ of the critical point
$r_{\text{hoop}}=2M$ above the new (and smaller) black-hole horizon
$r^{\text{new}}_{\text{H}}=2(M-{\cal E})$ is given by the integral
relation \cite{Bekt}
\begin{equation}\label{Eq6}
l_{\text{hoop}}({\cal E})=\int_{r^{\text{new}}_{\text{H}}=2(M-{\cal
E})}^{r_{\text{hoop}}=2M}{{dr}\over{\sqrt{1-{{2M}/{r}}}}}\ ,
\end{equation}
which yields
\begin{equation}\label{Eq7}
l_{\text{hoop}}({\cal E})=4\sqrt{M{\cal E}}\cdot[1+O({\cal E}/M)]\
.
\end{equation}

According to the Thorne hoop conjecture \cite{Thorne}, a necessary
condition for an Hawking field mode of proper wavelength
${\lambda({\cal E})}$ to be able to escape to infinity is provided
by the requirement \cite{Notesim}
\begin{equation}\label{Eq8}
\lambda({\cal E})\gtrsim l_{\text{hoop}}({\cal E})\  .
\end{equation}
The inequality (\ref{Eq8}) reflects the physical fact that, in order
{\it not} to be engulfed by an horizon, the emitted Hawking quantum
should avoid the situation of being entirely contained within the
critical hoop radius $r_{\text{hoop}}=2M$ \cite{Thorne}. Taking
cognizance of Eqs. (\ref{Eq5}), (\ref{Eq7}), and (\ref{Eq8}), one
immediately obtains the upper bound
\begin{equation}\label{Eq9}
{\cal E}\lesssim{\cal E}_{\text{max}}\equiv
{{\hbar^{2/3}}\over{M^{1/3}}}\
\end{equation}
on the energies of the emitted Hawking quanta which are allowed to escape the black hole to infinity.

\section{Summary and Discussion}

In the present compact paper we have analyzed the semi-classical
Hawking evaporation process of canonical Schwarzschild black holes.
Taking into account the non-linear nature of the interaction between
the Hawking quanta and the curved black-hole spacetime, it has been
shown that the proper frequencies of the emitted Hawking field modes
are bounded from above by the simple quantum relation [see Eq.
(\ref{Eq9})]
\begin{equation}\label{Eq10}
\omega\lesssim\omega_{\text{max}}=(\hbar M)^{-1/3}\  .
\end{equation}
In particular, we have emphasized the fact that, due to self-gravity
effects of the tunneling quanta, which manifest themselves through
the Thorne hoop conjecture \cite{Thorne}, higher energy Hawking
quanta are re-absorbed (rather than escape to infinity) by the black
hole.

Interestingly, the newly derived upper bound (\ref{Eq9}) on the
allowed energies of the emitted Hawking quanta is stronger than the
trivial bound (\ref{Eq4}), which follows from simple energy
conservation considerations, by the large dimensionless factor
$M^{4/3}/\hbar^{2/3}\gg1$ \cite{Notelrg}. In addition, it is
important to emphasize the fact that the maximally allowed energy
(\ref{Eq9}) of the emitted Hawking quanta is related to the
characteristic (average) energy (\ref{Eq3}) of the corresponding
black-hole emission spectrum by the large dimensionless ratio
 \begin{equation}\label{Eq11}
{{{\cal E}_{\text{max}}}\over{\bar{\cal E}}}\sim {{M^{2/3}}\over{\hbar^{1/3}}}\gg1\  .
\end{equation}
This relation implies that the non-linear self-gravity effect of the
tunneling quanta, which manifests itself through the Thorne hoop
conjecture [see Eq. (\ref{Eq8})], blocks only the large frequency
tail of the semi-classical Hawking emission spectrum.

\bigskip
\noindent
{\bf ACKNOWLEDGMENTS}
\bigskip

This research is supported by the Carmel Science Foundation.
I would like to thank Yael Oren, Arbel M. Ongo, Ayelet B. Lata, and Alona B. Tea for
stimulating discussions.



\begin{thebibliography}{99}
\bibitem{Notelm} The dimensionless parameters $\{l,m\}$ in the Hawking expression
(\ref{Eq1}) are the angular harmonic indices of the radiated field
modes.
\bibitem{Noteunit} We shall use natural units in which $G=c=k_{\text{B}}=1$.
\bibitem{Notesim} We use the $\sim$ symbol to denote an order-of-magnitude estimate.
\bibitem{Notelrg} We consider macroscopic black holes in the dimensionless $M^2/\hbar\gg1$ regime.
\end{thebibliography}
\end{document}